\documentclass[%
 reprint,
 amsmath,amssymb,
 aps,
]{revtex4-1}
\usepackage{physics}
\usepackage{graphicx}
\usepackage{dcolumn}
\usepackage{bm}
\usepackage{color}
\usepackage{hyperref}
\usepackage{float}

\usepackage{amssymb,amsmath,amsfonts}  
\usepackage[hang,scriptsize,tight]{subfigure}  

\begin{document}
\title{Noncommutative coherence and quantum phase estimation algorithm}

\author{Shubhalakshmi S$^{1,2}$}
\author{Ujjwal Sen$^{2}$}%
\affiliation{%
 $^{1}$Indian Institute of Science Education and Research, Pune 411 008, India 
}%

\affiliation{%
 $^{2}$Harish-Chandra Research Institute, HBNI, Chhatnag Road, Prayagraj 211 019, India 
}%




\date{\today}

\begin{abstract}
We present a measure of quantum coherence by employing the concept of noncommutativity of operators in quantum mechanics. We analyse the behaviour of this noncommutative coherence and underline its
similarities and differences with the conventional measures of quantum coherence. The maximally noncommutative coherent states turn out to be far removed from the conventionally considered maximally coherent states. We argue that the efficiency of the quantum phase estimation algorithm, an important rung in the Shor factorisation algorithm, is potentially related to the measure of noncommutative coherence.

\end{abstract}

\maketitle


\section{\label{sec:level1}Introduction}




The superposition principle within the quantum physics description of nature marks a signature departure from the classical physics description of the same. There has been significant effort towards understanding its counter-intuitive nature. One of the effects of the superposition principle is the phenomenon of entanglement, and has had ramifications in diverse areas ranging from foundations to technology \cite{HHHH,STUA}. The phenomenon of quantum coherence, also a product of the same principle, is as yet a fledgling area of research \cite{Aberg,BCP}.

Coherence has been of research interest since the advent of wave theory. It is a concept that is pivotal to the interference phenomenon, and has applications beyond ray optics and the classical regime. Quantum mechanics, known for unification of wave and particle natures has further strengthened the role of coherence in physics. In quantum information-theoretic terms, quantum coherence is defined as the entity that quantifies the ``amount of superposition" using a fixed reference basis. Recently, this has helped develop a resource-theoretic interpretation of quantum coherence \cite{Aberg,BCP,WY,MPP,BKW,QLP,CB,BBP,HS,CPPM,MBMP,MAL,SSDBA,ACR,ZYH,AESMAM,XTF,S,HZZSV,HML,NFM,YXLC,JYGVM,
YS,MMJM,PJ,KMJD,H,NA,HSXWZH,GBF,ZHC,ACFN,MJA,JTD,DC,JZJH,MG,SIAU,TDLM,KTGCGMA,XAHSUJ,XZYS,JA,EG,
ASMM,TNDM,SCSSAU, SZ}. 

The concept of noncommutativity of operators in quantum mechanics, ultimately attributable again to the superposition principle, forms one of the cornerstones of the theory, and appears in a broad spectrum
of aspects of the theory from the uncertainty principle \cite{WH,EH,HW,HPR,ESP,ESS} and quantum teleportation \cite{XX,LLV,SGAU} to out-of-time-ordered correlators \cite{Y}. Here we define a measure of quantum coherence by dissecting the noncommutativity properties of density operators in quantum mechanics, while keeping a fixed reference basis. We try to analyse the functional behaviour numerically and also find out its similarities and differences with respect to conventionally defined coherence measures. We find that the most coherent qubit state in case of noncommutative coherence is very close to the one making an angle of $\frac{\pi}{5}$ with the north pole on the Bloch sphere, while the state corresponding to $\frac{\pi}{2}$ which corresponds to the maximally coherent states for conventional quantum coherences, is at a local minimum. We try to study the behaviour of the function across mixed states and pure states. We also consider variations of the functional that defines the noncommutative coherence, resorting to different integer and fractional powers of the density matrices involved. Along with knowing the functional characteristics, we tried to uncover whether they have anything to say about a quantum-enabled protocol. We find that the efficiency in non-Hadamard versions of the quantum phase estimation algorithm, a crucial element of the Shor factorisation algorithm \cite{Shor}, is related to the noncommutative coherence  of the eigenbases of the non-Hadamard operations.

This paper is further divided into sections explaining specifics. In section \ref{sec:level2}, we present the motivations behind and the definition of noncommutative coherence. Section \ref{sec:level3} deals with the computational results for understanding the function defined in the preceding section. Section \ref{sec:level4} explains how the details of the functional extrema relates to the success probability of the quantum phase estimation algorithm with non-Hadamard gates. A conclusion is presented in section \ref{sec:level11}.

\section{\label{sec:level2}Motivations and Definition}

In quantum mechanics, the state of a physical system is described by using a density matrix. An interesting aspect of density matrices is that they may not mutually commute. Here, we try to construct and physically interpret a mathematical definition of quantum coherence inspired by the properties of noncommuting matrices. 

We begin with the observation that for two density matrices, $\rho$ and $\sigma$, if they commute, then $(\rho \sigma)^{\frac{1}{2}}$ is the same as $\rho^{\frac{1}{2}} \sigma^{\frac{1}{2}}$. The fact holds for any other power as well. It of course trivially holds for the unit power, even for noncommuting pairs. We will later on consider powers that are  non-unit and also not $\frac{1}{2}$. But we begin, for specificity, with the power $\frac{1}{2}$. A density operator is hermitian, positive semi-definite and of unit trace. If  $\rho$ and $\sigma$ do not commute, then $\rho \sigma$ may not even be hermitian. Just like for observables in quantum mechanics, this can be remedied by considering the symmetrised version, viz. $\rho \sigma + \sigma \rho$. For commuting density matrices, $(\frac{1}{2}(\rho \sigma + \sigma \rho))^\frac{1}{2}$ and $\frac{1}{2}(\sqrt{\rho}\sqrt{\sigma} + \sqrt{\sigma}\sqrt{\rho})$ are again equal. They are not equal if $\rho$ and $\sigma$ are noncommuting. In fact, $\rho \sigma + \sigma \rho$ and $\sqrt{\rho} \sqrt{\sigma} + \sqrt{\sigma} \sqrt{\rho}$, are although always hermitian, may not always be positive semi-definite. We deal with this problem by considering the operators $(\frac{1}{2} |\rho \sigma + \sigma \rho|)^\frac{1}{2}$ and $\frac{1}{2}|\sqrt{\rho}\sqrt{\sigma} + \sqrt{\sigma}\sqrt{\rho}|$. Again, they are equal if $\rho$ and $\sigma$ commute.

The notion of quantum coherence usually begins with a naturally-defined set of distinguishable, i.e., orthogonal pure states of the system under consideration. In case, e.g., of an interferometric experiment, this is formed by the states representing the arms of the interferometer. The states in that set of orthogonal states, forming a basis of the space spanned by them, and their mixtures are then considered as states with vanishing quantum coherence. The qualitative concept of quantum coherence stems from this premise, viz., any state that is not a mixture of the chosen orthonormal basis, i.e., any state that remains non-diagonal when written in the chosen orthonormal basis, is coherent. We are ``disregarding" here the set of works considering quantum coherence with respect to a non-orthogonal basis \cite{TNDM,SCSSAU}. The quantitative path is however very diverse \cite{Aberg,BCP,WY,MPP,BKW,QLP,CB,BBP,HS,CPPM,MBMP,MAL,SSDBA,ACR,ZYH,AESMAM,XTF,S,HZZSV,HML,NFM,YXLC,JYGVM,
YS,MMJM,PJ,KMJD,H,NA,HSXWZH,GBF,ZHC,ACFN,MJA,JTD,DC,JZJH,MG,SIAU,TDLM,KTGCGMA,XAHSUJ,XZYS,JA,EG,
ASMM,TNDM,SCSSAU, SZ}, but in some way or the other tries to measure the non-diagonal terms, in the chosen
basis..

In this paper, we go beyond this narrative via measuring the quantum coherence of a state $\rho$ by measuring its noncommutativity with arbitrary states $\sigma$, with the latter being chosen as mixtures in the chosen basis. The noncommutativity between $\rho$ and $\sigma$ is in turn quantified by the distance between the operators $(\frac{1}{2} |\rho \sigma + \sigma \rho|)^\frac{1}{2}$ and $\frac{1}{2}|\sqrt{\rho}\sqrt{\sigma} + \sqrt{\sigma}\sqrt{\rho}|$, by using a suitable distance measure.

Deferring the choice of the distance measure, we are still left with the problem of choosing a suitable incoherent state $\sigma$. The distance between two surfaces can be quantified in a multitude of ways. One of them is to consider the point, in the parameter space that generates the surfaces, where they reach the closest. An antipodal concept is to consider the point where they remain the farthest. One can also consider several intermediate ones. Likewise, for a fixed $\rho$, we have two ``surfaces" at hand, viz., $(\frac{1}{2} |\rho \sigma + \sigma \rho|)^\frac{1}{2}$ and $\frac{1}{2}|\sqrt{\rho}\sqrt{\sigma} + \sqrt{\sigma}\sqrt{\rho}|$, with the generating ``parameter" being the set of incoherent states $\sigma$. Let us first consider the \textit{minimum} parameter point approach, so that the corresponding quantity is $\underset{\sigma \in \textit{I}} \min  \; \mathcal{D}\left[ (\frac{1}{2} | \rho \sigma + \sigma \rho |)^{\frac{1}{2}} , \frac{1}{2} |\sqrt{\rho} \sqrt{\sigma} + \sqrt{\sigma} \sqrt{\rho} | \right] $, where  \textit{I} denotes the set of incoherent states, and $\mathcal{D}$ denotes a distance function.

Choosing the above definition, we are led to the following roadblock.
The identity operator is an element of every possible set of incoherent states and commutes with every density operator defined on \(\mathbb{C}^d\)  implying the minimum distance to be always zero. Our ``surfaces" $(\frac{1}{2} |\rho \sigma + \sigma \rho|)^\frac{1}{2}$ and $\frac{1}{2}|\sqrt{\rho}\sqrt{\sigma} + \sqrt{\sigma}\sqrt{\rho}|$, therefore touch each other at least at one ``point", viz. for every $\rho$, we can choose the $\sigma$ to be the identity operator in the corresponding Hilbert space. There are of course many ways to remedy this difficulty, and we consider the strategy where the \textit{maximum} distance between the surfaces is deemed fit to quantify the distance between the surfaces.

The definition of the measure is therefore as follows:
\begin{equation}
C_{nc}(\rho)=\underset{\sigma \in \textit{I}} \max \; \mathcal{D}\left[ (\frac{1}{2} \mid \rho \sigma + \sigma \rho \mid)^{\frac{1}{2}} , \frac{1}{2} \mid \sqrt{\rho} \sqrt{\sigma} + \sqrt{\sigma} \sqrt{\rho} \mid \right] \label{eq:main}
.\end{equation}

Since the definition focuses on measuring the amount with which the two resultant operators differ, mainly by exploiting the noncommutative nature of the two density operators involved, we term the quantity as ``noncommutative coherence". Along with studying the properties of this quantity, we also study how it resembles vis-{\` a}-vis differs from the conventionally defined quantum coherence measures.

\section{\label{sec:level3}Traits of noncommutative coherence}
We now try to analyse the quantity defined in the preceding section by first considering the density operators, $\rho$, defined over $\mathbb{C}^{2}$. We choose the computational basis as the reference basis.
The incoherent states, $\sigma \in \textit{I}$, are therefore expressible in the form,
$$\sigma=
\begin{bmatrix}
p & 0 \\
0 & 1-p
\end{bmatrix}, $$
where $0 \leq p \leq 1$ and `$p$' is the parameter that is to be optimised. The matrix is expressed in computational basis.

A general density matrix, $\rho = \frac{1}{2} ( \mathbb{I}_{2} + \vec{r}. \vec{\sigma} )$, in matrix form, in the computational basis, can be written as 
$$\rho=
 \begin{bmatrix}
\frac{1+ r \cos\theta}{2} &   \frac{r \sin\theta}{2} \\
\\
 \frac{r \sin\theta}{2} & \frac{1- r \cos\theta}{2}
\end{bmatrix} , $$
where the $0 < r \leq 1$ and $0 < \theta < \pi$. Here we have not considered the phase term as it does not play a major role in our considerations. We have also removed the points in the parameter space for which $\rho$ becomes a member of $\textit{I}$. 
Here, $\mathbb{I}_2$ denotes the identity operator on the qubit Hilbert space, $\vec{\sigma}=(\sigma_{x},\sigma_{y},\sigma_{z})$ denotes the triad of Pauli spin-$\frac{1}{2}$ matrices, and $\vec{r}$ is the Bloch vector. 
And $\theta$ is the zenith angle (angle from the direction $\theta=0$, which represents the quantum state $\ket{0}$) on the Bloch Sphere. 
$r\equiv |\vec{r}|$ is $\leq1$ in general with $r=1$ being the case of pure states. Of course, $r\geq 0$ for all states.

We now consider relative entropy \cite{Wehrl} as our distance function. The relative entropy distance of $\rho$ from $\sigma$ is given by 
\begin{equation}
S(\rho || \sigma) = \mbox{tr}(\rho \log_{2} \rho- \rho \log_{2} \sigma) .
\end{equation}
It is to be noted that the relative entropy ``distance" is not symmetric and also does not satisfy the triangle inequality. 

\subsection{When $\rho$ is pure}
\begin{figure}
\center
\includegraphics[width=3.5in]{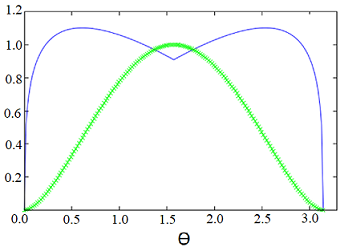} 
  \caption{Noncommutative coherence for pure qubits. The continuous blue line represents the noncommutative coherence of the state $\cos\frac{\theta}{2}\ket{0} + \sin\frac{\theta}{2}\ket{1}$, with $\left\lbrace \ket{0},\ket{1} \right\rbrace$ forming the computational basis. The profile with green crosses represents the relative entropy of coherence in the same basis. While the horizontal axis is for $\theta$ and is measured in radians, the vertical axis is for noncommutative coherence for the continuous line and for  relative entropy of coherence for the crosses. In both cases, the vertical line is measured in bits.}
  \label{Fig.1:Pure State}
\end{figure}

In case the state of the qubit is pure, it can be expressed as
$$\rho=
\begin{bmatrix}
\cos^2\frac{\theta}{2} & \frac{\sin\theta}{2} \\
\\
\frac{\sin\theta}{2} & \sin^2\frac{\theta}{2}
\end{bmatrix}, $$
ignoring the phase, that is not relevant for our purposes. For a pure state, we have $ \sqrt{\rho} = \rho$. Eq. (\ref{eq:main}) now simplifies to
\begin{equation}
C_{nc}(\rho)= \underset{\sigma \in \textit{I}} \max \; D\left[ (\frac{1}{2}\mid \rho \sigma + \sigma \rho \mid)^{\frac{1}{2}} , \frac{1}{2}\mid \rho \sqrt{\sigma} + \sqrt{\sigma} \rho \mid \right]
.\end{equation}
We numerically compute $C_{nc}$ for all pure $\rho$ and plot the results in Fig. \ref{Fig.1:Pure State}, and also plot the conventional quantum coherence for comparison. We use the relative entropy of coherence \cite{Aberg, BCP} as the measure of conventional quantum coherence, being defined for $\rho$ as
\begin{equation}
C(\rho)= \underset{\sigma \in \textit{I}}\min \; S(\rho || \sigma)
.\end{equation}
The results obtained and the corresponding comparisons remain qualitatively unaltered by replacing $C(\rho)$ with the trace distance-based quantum coherence. We find that noncommutative coherence behaves quite differently from the conventional quantum coherence. Of course, they share the property of vanishing at the poles of the Bloch sphere, i.e., for $\theta=0,\pi$. The conventional coherence measure attains its maximum at $ \theta = \frac{\pi}{2}$. But in case of the noncommutative coherence, although $ \theta = \frac{\pi}{2}$ is still a special point, it does not provide the maximum $C_{nc}$. $ \theta = \frac{\pi}{2}$ is still a ``special" point for $C_{nc}$ in the sense that it provides a local minimum in the parameter space and is the only point of non-analyticity. Also, and like $C$, the profile of  $C_{nc}$ as a function of $\theta$ is symmetric about $ \theta = \frac{\pi}{2}$. But, in direct contradistinction to the conventional quantum coherence measure, noncommutative coherence has two maxima which are symmetric about $ \theta = \frac{\pi}{2}$, being at a finite distance from the latter.

\subsection{When $\rho$ is mixed} 
\begin{figure}
\center
\includegraphics[width=3.5in]{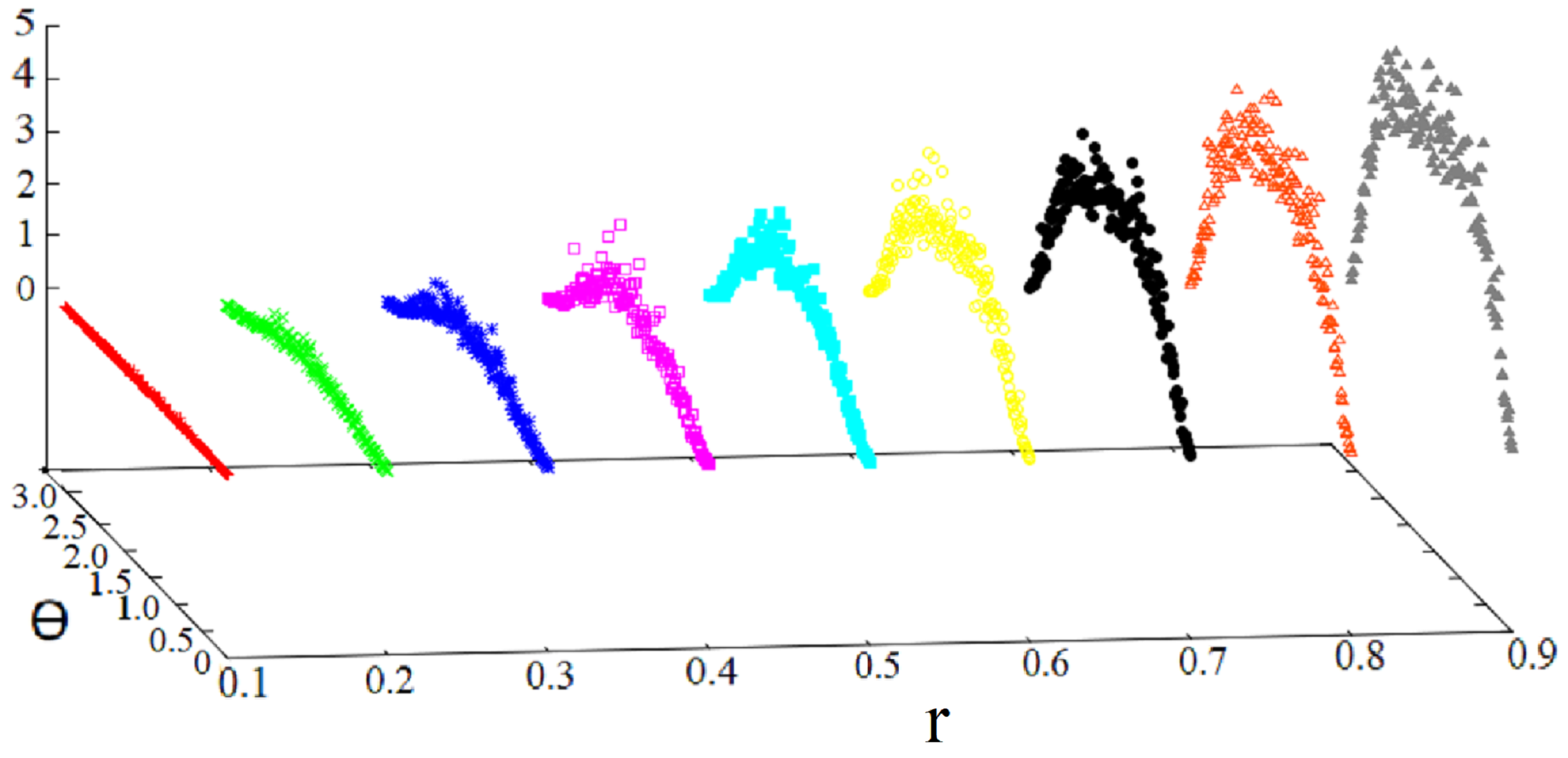} 
  \caption{ Noncommutative coherence for general qubits. The base represents the parameter space of an arbitrary qubit, with $r$ representing the radius of the corresponding Bloch vector and $\theta$ the zenith angle. The azimuthal angle is not relevant for the case at hand. The vertical axis represents the noncommutative coherence, $C_{nc}$, of a qubit state corresponding to the parameter values on the base. The base axes are dimensionless while the vertical one is in bits. It may be unsettling to find that while the noncommutative coherence for pure qubits attains a value close to unity, the same for mixed states with \(r=0.9\) reaches values close to 5. The values however again go down continuously from about 5 to about unity as we go from \(r=0.9\) to 
  \(r=1\), as seen in Fig. \ref{Fig.2b:Mixed State 2}.}
  \label{Fig.2a:Mixed State}
\end{figure}

\begin{figure}
\center
\includegraphics[width=3.5in]{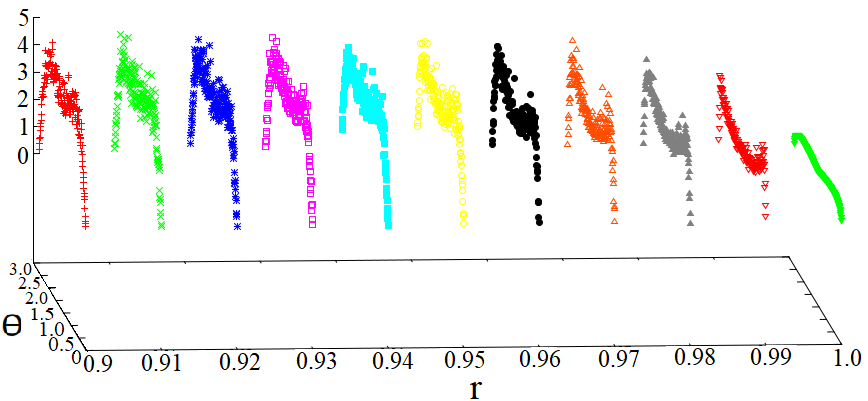} 
  \caption{ Noncommutative coherence for nearly-pure nonpure qubits. All considerations in this figure remains the same as in the preceding figure, except that we focus on the range \(r \in [0.9, 1]\). 
  }
  \label{Fig.2b:Mixed State 2}
\end{figure}

We now  move over to the case of a mixed qubit, which can be expressed as $\rho = \frac{1}{2} ( \mathbb{I}_2 + \vec{r}. \vec{\sigma} )$, where 
\(0<r<1\). We have ignored the \(r=0\) case as it corresponds to an incoherent state, and the case for which \(r=1\) as that has already been considered above. 
The  behaviour of noncommutative coherence for a mixed state $\rho$, with fixed (non-unit, non-zero) $r$, is similar to the previous case where $\rho$ was taken to be pure (\(r=1\)). There is again a local minima approximately at  $ \theta = \frac{\pi}{2}$, 
but we observe that the dip of $C_{nc}$ at the local minima becomes less prominent as $r$ decreases from unity to zero. See Figs. 
\ref{Fig.2a:Mixed State} and \ref{Fig.2b:Mixed State 2}.

\subsection{Varying powers} 
In this subsection, we try to understand the behaviour of noncommutative coherence by considering the other powers of the density operators and not just the square root. We consider only pure $\rho$ for the function examination. In general, the definition of noncommutative coherence of order $n$ is given by
\begin{equation}
C^{(n)}_{nc}(\rho)=\underset{\sigma \in \textit{I}} \max \; \mathcal{D}\left[ (\frac{1}{2} \mid \rho \sigma + \sigma \rho \mid)^{\frac{1}{n}} , \frac{1}{2} \mid \rho^{\frac{1}{n}} \sigma^{\frac{1}{n}} + \sigma^{\frac{1}{n}} \rho^{\frac{1}{n}} \mid \right] \label{eq:main2}
.\end{equation} 
 
\begin{figure}
\center
\includegraphics[width=3.5in]{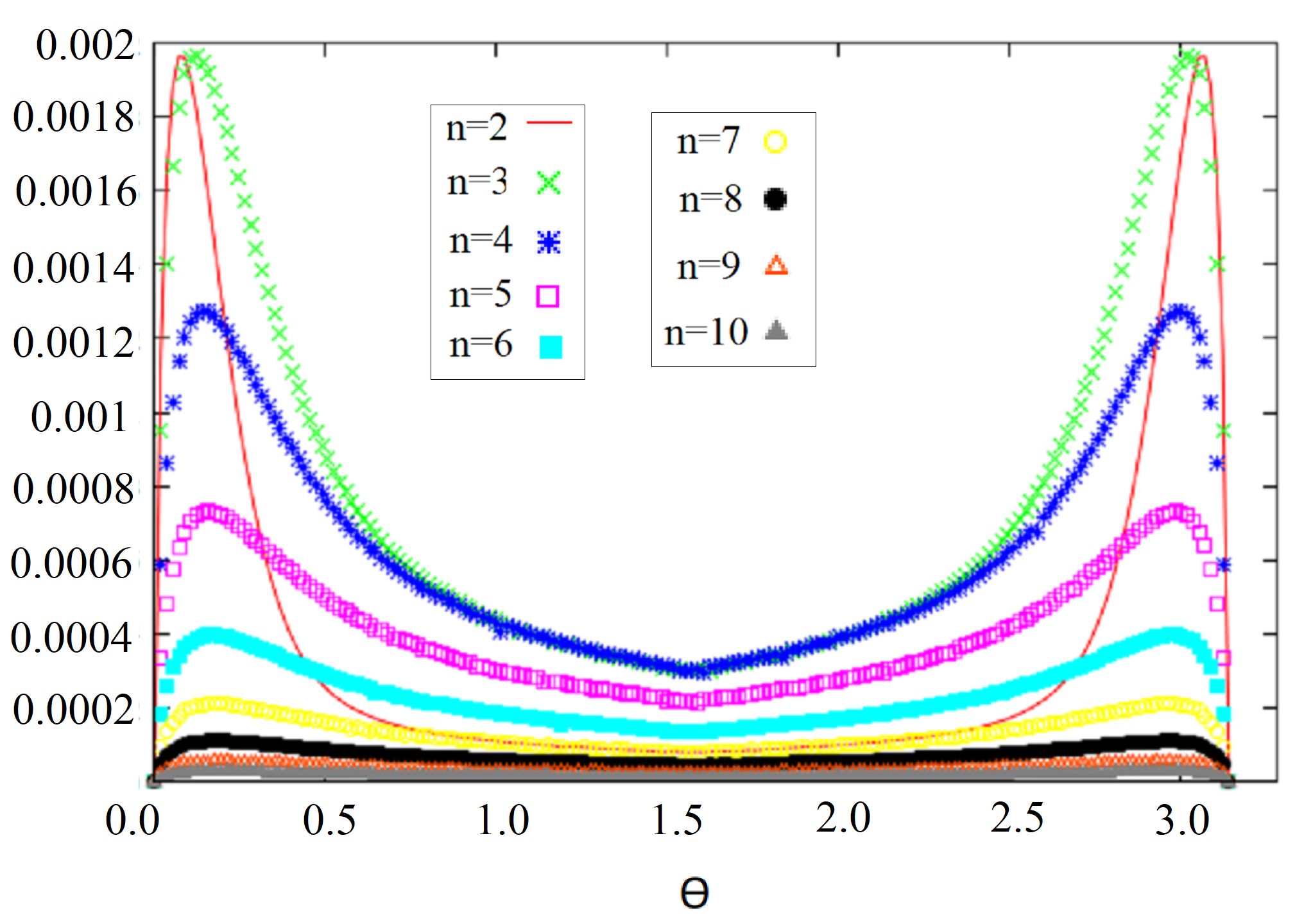} 
  \caption{ Noncommutative coherences of different integer orders for pure states. We plot here noncommutative coherences of different orders on the vertical axis against the state parameter $\theta$ on the horizontal one. For a given $\theta$, the noncommutative coherences are evaluated for the state $\cos\frac{\theta}{2}\ket{0} + \sin\frac{\theta}{2}\ket{1}$, with $\left\lbrace \ket{0},\ket{1} \right\rbrace $ forming the computational basis. The values of the orders of the noncommutative coherences can be read off from the plots by noting the symbols in the legend. While the vertical axis is in bits, the horizontal one is in radians.}
  \label{Fig.3:Powers in integers}
\end{figure}

\begin{figure}
\center
\includegraphics[width=3.5in]{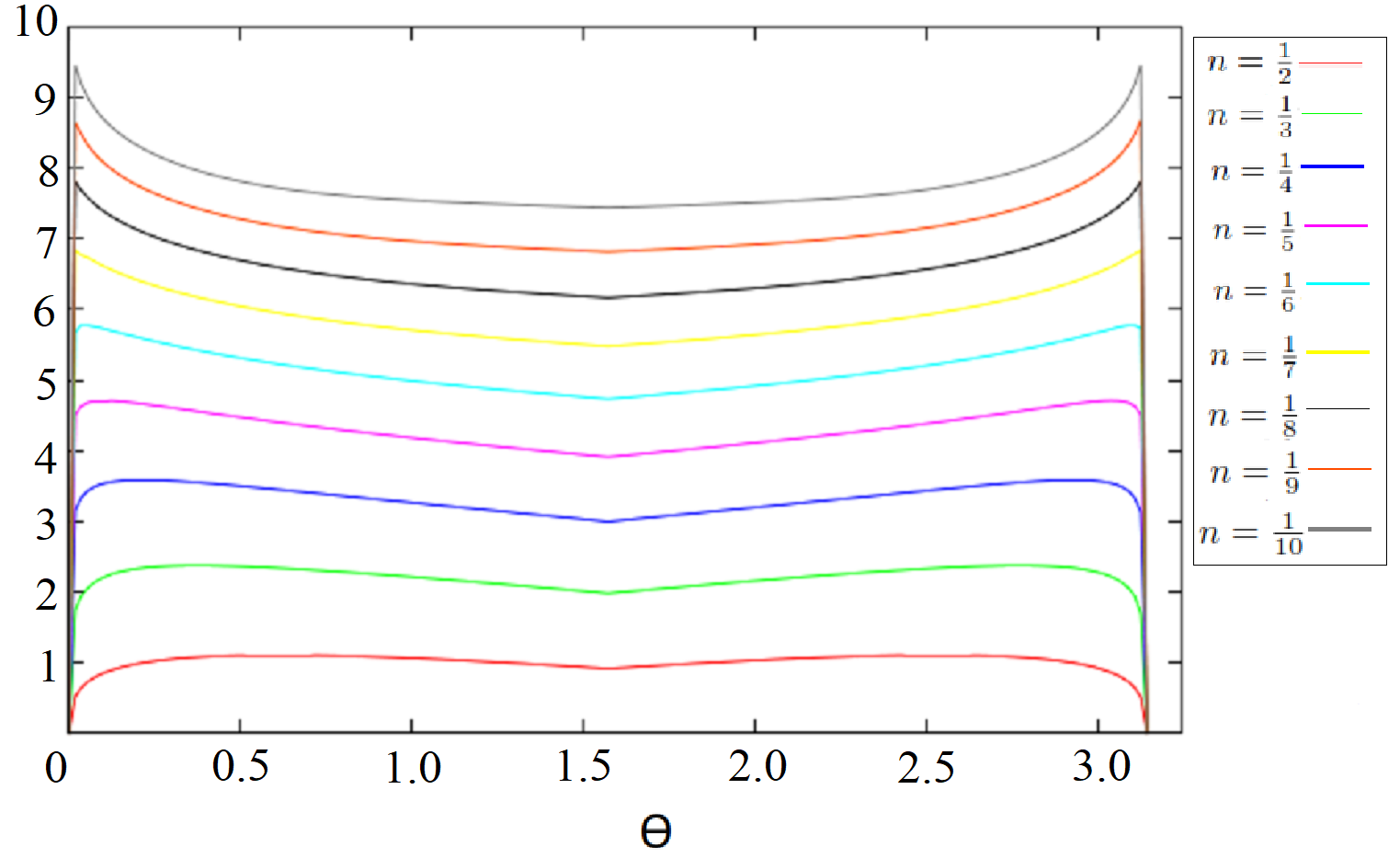} 
  \caption{ Noncommutative coherences of different fractional orders for pure states. The input conditions are just as in Fig. \ref{Fig.3:Powers in integers}, except that now $n$ runs through $\frac{1}{2}, \frac{1}{3},.., \frac{1}{10}.$ At the middle of the horizontal axis, their values successively increase as we go from $\frac{1}{2}$ to $\frac{1}{10}$.}
  \label{Fig.4:Powers in fractions}
\end{figure}

We subdivide the types of values that $n$ can take, into cases when it is an integer and when the same is fractional. The ``usual" noncommutative coherence is then the noncommutative coherence of order 2. The behaviour of noncommutative coherences of different orders is portrayed in Figs. \ref{Fig.3:Powers in integers} and \ref{Fig.4:Powers in fractions}. The noncommutative coherence of order unity is of course zero for all quantum states.

\paragraph*{n taking integer values.}
From the plots in Fig. \ref{Fig.3:Powers in integers}, we observe that the local minima are always at $\theta = \frac{\pi}{2}$. The two maxima, symmetric about the corresponding local minimum, still exist, and they become sharper and move towards poles as $n$ increases.

\paragraph*{n taking fractional values.}
From the plots in Fig. \ref{Fig.4:Powers in fractions}, we again observe two distinct maxima. As $\frac{1}{n}$ increases, 
the maxima move closer to the poles. 

\section{\label{sec:level4} Quantum phase estimation algorithm: Non-Hadamard versions}
The notion of noncommutative coherence is conceptualized for quantifying the physical phenomenon of quantum coherence. There are several measures of quantum coherence, but they usually agree on certain broad contours \cite{Aberg, BCP,WY}. One of them is that the state $\frac{1}{\sqrt{2}}(\ket{0} + \ket{1})$ is a state having the maximum quantum coherence, among states of a qubit. This fits well with the fact that $\left\lbrace \ket{0}, \ket{1} \right\rbrace $ and $\lbrace \frac{1}{\sqrt{2}}(\ket{0} \pm \ket{1})\rbrace$ form mutually unbiased bases \cite{IDI,WBD,SPVF} of the qubit Hilbert space. Arguably, it is with a similar intuition that these two mutually unbiased bases are used in Bennett-Brassard 1984 quantum key distribution protocol \cite{BB84}.

The notion of noncommutative coherence moves against this intuition by settling $\frac{1}{\sqrt{2}}(\ket{0} + \ket{1})$ with a non-maximal noncommutative coherence. One is reminded here of similar situations in entanglement theory, where nonmaximally entangled states may either behave equal or ``superior" to maximally entangled states \cite{HMPEPC,ADGL,HAU,ASAU,SASAU,CSAU}, in certain quantum-enabled protocols. It is in quest of a similar quantum-enabled protocol that provides operational meaning to the state with maximal noncommutative coherence that we analyse the quantum phase estimation algorithm in its non-Hadamardian avatars.

\begin{figure}
\center
\includegraphics[width=3.5in]{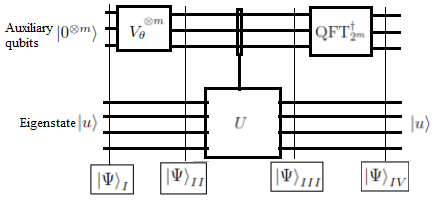} 
  \caption{ A schematic diagram of the quantum phase estimation circuit. See text for details.}
  \label{Fig.5:PEA}
\end{figure}

The quantum phase estimation algorithm (QPEA) is one of the most basic quantum algorithms which has a wide range of applications \cite{KSV,JW}. As the name suggests, it is an algorithm that approximately determines the phase introduced by a unitary on a specific eigenstate of the unitary. Let the unitary be a p-qubit one and be denoted by U. Let the specific eigenstate be denoted by $\ket{u}$. In the standard QPEA, first an $m$-qubit auxiliary is passed through m Hadamard gates and the resultant states act as states of control qubits for a controlled unitary operator to act in the next stage. The eigenstate is taken to be the state of the target qubits for the controlled unitary. Phase kick-back is observed on auxiliary qubits, and the target qubits are discarded. Finally, the inverse quantum Fourier transform is performed on the auxiliary qubits, and a subsequent measurement onto the computational basis results in the phase estimation. The precision of the estimation depends on the number of auxiliary qubits. See Fig. \ref{Fig.5:PEA}, the circuit in which corresponds to the usual quantum phase estimation if the unitary operator $V$ in the circuit is replaced by the Hadamard operator.

Intuitively, the success of the quantum phase estimation algorithm lies in the noncommutativity between operators that are diagonal in the computational basis ($m$-fold tensor products of the states in $\left\lbrace \ket{0},\ket{1}\right\rbrace $) and those that are diagonal in  the $H^{\bigotimes m}$-rotated one ($m$-fold tensor products of the states in $\lbrace  \frac{1}{\sqrt{2}} (\ket{0}\pm \ket{1})\rbrace $), that is incorporated in the $m$ auxiliary qubits by the Hadamard gates. Here, $H$ denotes a Hadamard gate. The controlled unitary operator subsequently exploits the idea of quantum parallelism on the huge superposition, in the computational basis, thus created. The unusual behaviour of noncommutative coherence motivates us to look at a variant of the quantum phase estimation algorithm. More precisely, the fact that noncommutative coherence provides a higher value for a state that is not $\frac{1}{\sqrt{2}} (\ket{0} + \ket{1})$, even after subsuming the phases in the definitions of $\ket{0}$ and $\ket{1}$, motivates us to look at the efficiencies of variants of the quantum phase estimation algorithm that uses non-Hadamard gates for exploitation of the resulting state by a subsequent quantum parallelism technique.

We replace the Hadamard gate in the standard phase estimation circuit with a general unitary, in place of $H$ that brings about a $\theta$-rotation to the auxiliary qubit $\ket{0}$, in the plane defined by the bases $\left\lbrace \ket{0}, \ket{1} \right\rbrace $ and $\lbrace  \frac{1}{\sqrt{2}} (\ket{0}\pm \ket{1})\rbrace $ of the Bloch ball. We study how the success probability of the algorithm varies with respect to the angle of rotation, $\theta$.
The unitary gate is of the form 
$$V_{\theta}=
\begin{bmatrix}
\cos\frac{\theta}{2} & \sin\frac{\theta}{2} \\
\\
\sin\frac{\theta}{2} & -\cos\frac{\theta}{2}
\end{bmatrix} $$

The quantum circuit for the algorithm is schematically presented in Fig. \ref{Fig.5:PEA}.
Let us briefly go through the steps of the circuit. The input to the circuit is an $m$-qubit state $\ket{0}^{\otimes m}$, and a $p$-qubit state $\ket{u}$, with the latter being a specific eigenstate  of a $p$-qubit unitary $U$. Being a unitary, the eigenvalues of $U$ are phases, and the task of the quantum phase estimation algorithm is to estimate this phase for a given $U$ and a given eigenstate $\ket{u}$ of the unitary $U$. Let us denote the $(m+p)$-qubit input as $\ket{\Psi}_{I}$, so that $\ket{\Psi}_{I} = \ket{0}^{\otimes m} \; \otimes \; \ket{u}$. We then apply $V_{\theta}^{\otimes m}$ on the first $m$ qubits to obtain 
\begin{equation}
\ket{\Psi}_{II} = (\cos\frac{\theta}{2}\ket{0} + \sin\frac{\theta}{2})^{\otimes m} \; \otimes \; \ket{u}. \\
\end{equation}
We then apply a controlled-unitary operator, defined as follows, on the $(m+p)$-qubit state $\ket{\Psi}_{II}$: $\ket{K} \ket{\phi} \rightarrow \ket{K} (U^{K}\ket{\phi}) $,
where $K = 0,1,2,.., 2^{m}-1, \ket{K} \in (\mathbb{C}^{2})^{\otimes m}$ and $\ket{\phi}$ is an arbitrary vector in $(\mathbb{C}^{2})^{\otimes p}$. Let us denote the output of the controlled-unitary as $\ket{\Psi}_{III}$. Also, set $U\ket{u}= e^{2 \pi i \phi}\ket{u}$, where $\phi$ is a real number. Then
\begin{equation}
\ket{\Psi}_{III} = \sum_{y=0}^{2^{m}-1} (\cos\frac{\theta}{2})^{m-b(y)} (\sin\frac{\theta}{2})^{b(y)} e^{2 \pi i \phi y} \ket{y} \otimes \ket{u},\\
\end{equation}
where $b(y)$ is the number of 1's in the binary decomposition of the number $y$, with $y=0,1,2,..,2^{m}-1$.\\ Note here that 
\begin{equation*}
V_{\theta} \ket{0} = \cos\frac{\theta}{2}\ket{0} + \sin\frac{\theta}{2}\ket{1}.
\end{equation*}
We now apply the hermitian adjoint of the quantum Fourier transform on $\ket{\Psi}_{III}$, where the quantum Fourier transform acts as 
\begin{equation*}
\ket{j} \rightarrow \frac{1}{\sqrt{2^{m}}} \sum_{K=0}^{2^{m}-1} e^{\frac{2 \pi i j k}{2^{m}}}\ket{K},
\end{equation*}
with $j=0,1,2,..,2^{m}-1$.\\ The output, denoted as $\ket{\Psi}_{IV}$, is given by
\begin{equation}
\ket{\Psi}_{IV} = \frac{(\cos\frac{\theta}{2})^{m}}{2^{\frac{m}{2}}} \sum_{a=0}^{2^{m}-1} \sum_{y=0}^{2^{m}-1} (\tan\frac{\theta}{2})^{b(y)} e^{i \frac{2 \pi (a-x) y}{2^{m}}} e^{2 \pi i \delta y} \ket{x},
\end{equation}
where we here set $\phi = \frac{a}{2^{m}} + \delta$ with $|\delta| \leq 2^{-(m+1)}$, for some $a=0,1,2,..,2^{m}-1$, and where we have ignored the last $p$ qubits, which are in the state $\ket{u}$.

Similar to the standard phase estimation algorithm, we have to perform a projective measurement of the first $m$ qubits in the computational basis. The probability that $\ket{a}$ ``clicks" in the measurement is given by

\begin{equation}
p_{a}= \mid \frac{(\cos\frac{\theta}{2})^{m}}{2^{\frac{m}{2}}} \sum_{y=0}^{2^{m}-1} (\tan\frac{\theta}{2})^{b(y)} e^{i2 \pi \delta y} \mid^{2}
.\end{equation}

\begin{figure}
\center
\includegraphics[width=3.5in]{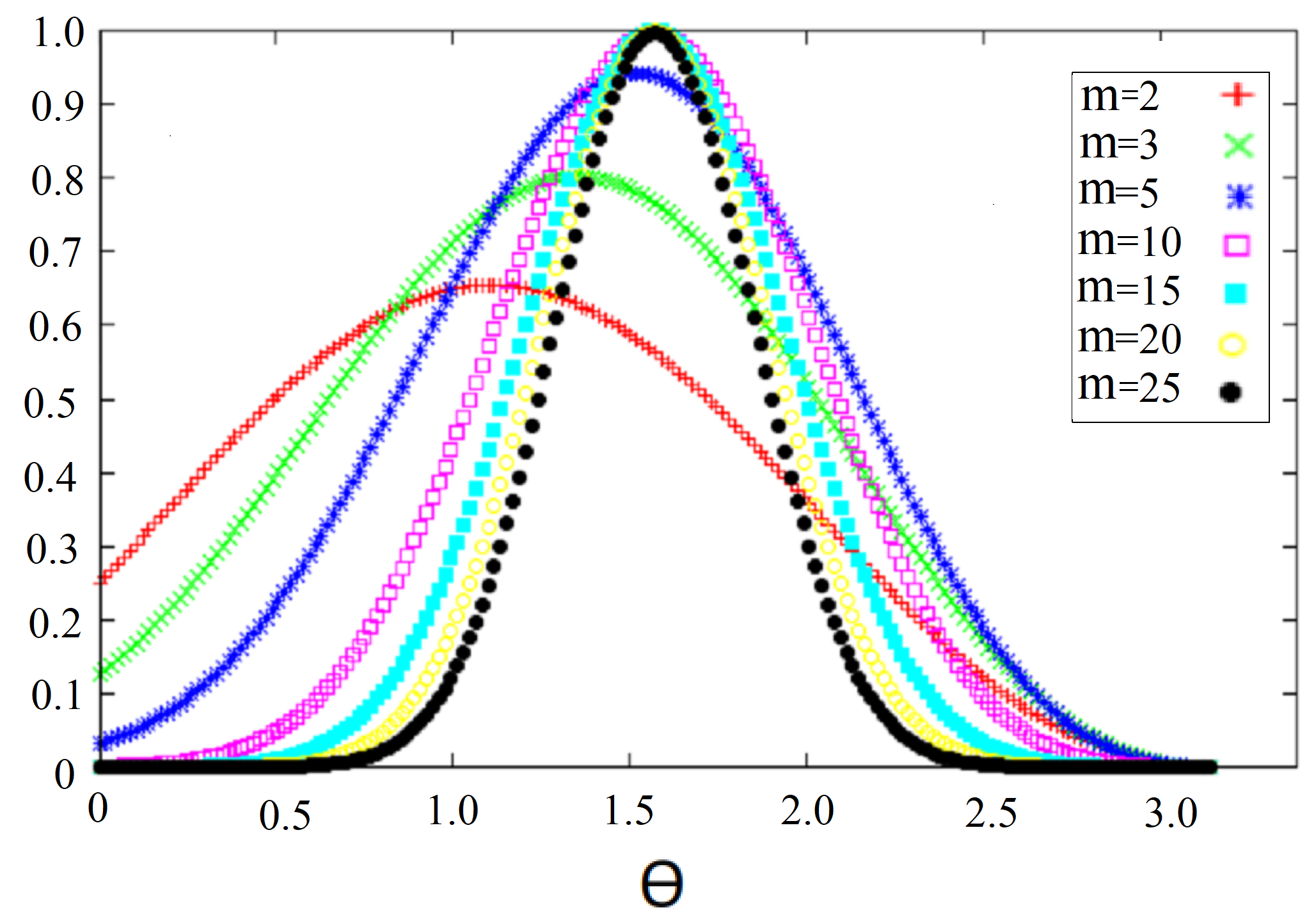} 
  \caption{Success probabilities in non-Hadamard quantum phase estimation algorithms. The horizontal axis represents $\theta$ in radians, as in the unitary $V_{\theta}$. The corresponding success probability is represented on the vertical axis. The vertical axis is dimensionless. The different plots are for different numbers of  auxiliary qubits, $m$, used in the algorithm. $\theta=\frac{\pi}{2}$ corresponds to the standard QPEA.}
  \label{Fig.6:SP}
\end{figure}

This is the probability that for a given $m, \theta, U, \ket{u}$, we are able to correctly predict the corresponding phase correct to $m$ bits. We plot this probability as a function of $\theta$ in Fig. \ref{Fig.6:SP}. For small values of $m$, the minimal probability is reached at different $\theta < \frac{\pi}{2}$.

\begin{figure}[H]
\center
\includegraphics[width=3.5in]{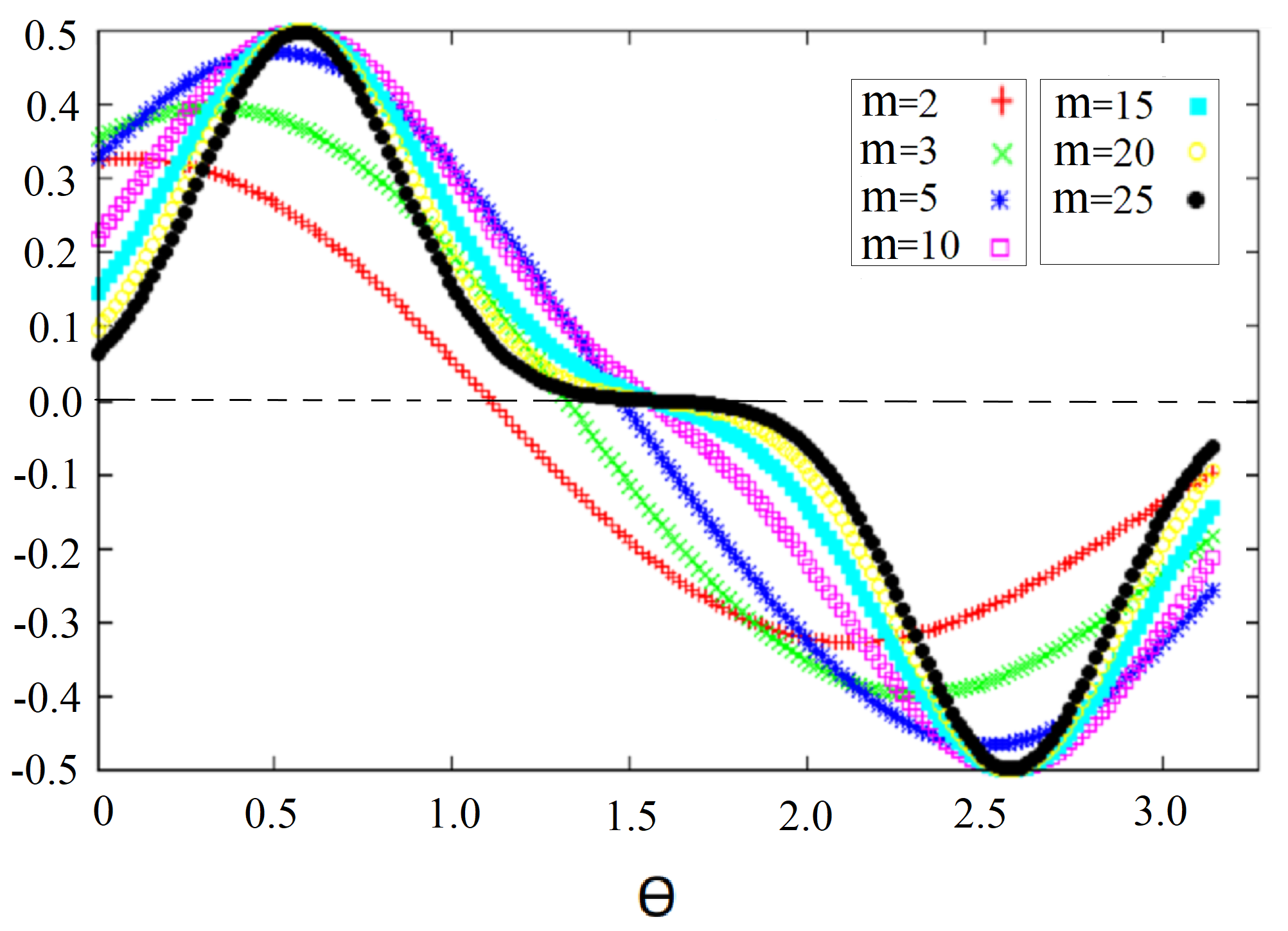} 
  \caption{Non-Hadamard quantum phase estimation algorithms and noncommutative coherence. The considerations and the symbols are the same as in Fig. \ref{Fig.6:SP}, except that the vertical axis represents the derivatives of $p_{a}$ with respect to $\theta$. For $m \geq 10$, the maximum of the derivative is approximately at $\theta=\frac{\pi}{5}$. Compare with Fig. \ref{Fig.1:Pure State}.}
  \label{Fig.7:DSP}
\end{figure}

However, as $m$ increases, the point of maximum gets closer to $\theta = \frac{\pi}{2}$ corresponding to which $V$ is the Hadamard gate, and we correspondingly have the standard quantum phase estimation algorithm. For the purpose of the plots, in Fig. \ref{Fig.6:SP}, as well as in Fig. \ref{Fig.7:DSP} described below, we have chosen $\delta = \frac{1}{2^{10}}$ for $2 \leq m \leq 7$,  $\delta = \frac{1}{2^{20}}$ for $7 < m \leq 17$, and  $\delta = \frac{1}{2^{30}}$ for $17 < m \leq 25$.

A careful look at the curves for $p_{a}$ as functions of $\theta$ reveals that in the regime of $\theta$ before $p_{a}$ reaches its maximum, there is a value of $\theta$ at which the probability changes its nature and starts to increase with $\theta$  at a significantly faster pace than for lower $\theta$. More precisely, there is a $\theta$ at which $p_{a}$ changes from being concave to being convex. This is more clearly seen in Fig. \ref{Fig.7:DSP}, where we plot the derivatives of $p_{a}$ with respect to $\theta$. For moderately  high values of $m$, e.g. for $m \gtrsim 10$, this change of curvature occurs at approximately $\theta= \frac{\pi}{5}$. Interestingly, this is also approximately the value of $\theta$ for which the noncommutative coherence is a maximum (see Fig. \ref{Fig.1:Pure State}) for the state $\cos\frac{\theta}{2}\ket{0} + \sin\frac{\theta}{2}\ket{1}$, the latter being precisely $V_{\theta}\ket{0}$, where $V_{\theta}$ is the unitary used in  the non-Hadamard version of the QPEA (see Fig. \ref{Fig.5:PEA}).

\section{\label{sec:level11}Conclusion}
While the concept of quantum coherence was well-known since the beginnings of quantum mechanics, it is only recently that a systematic quantification and resource-theoretic analysis of it has been performed. Just like the parallel, and earlier and arguably more well-established, resource theory of quantum entanglement, the resource theory of quantum coherence can be viewed from a variety of perspectives, and the connections and relative importance between them are not always very clear. In this paper, we looked at the concept of noncommutativity of operators in quantum mechanics to formulate a quantification of quantum coherence. We termed it as noncommutative coherence, and tried to analyse some of its properties. Similar to the conventional quantum coherence measures, the noncommutative coherence also depends on a chosen basis. An interesting fact about noncommutative coherence, with respect to the computational basis, is that the maximal noncommutative coherent pure qubit is not an equatorial state on the Bloch sphere, but is situated on a significantly higher latitude, approximately at \(\frac{3 \pi}{10}\). This
corresponds to a polar angle (or zenith angle) of approximately \(\frac{\pi}{5}\).

The quantum phase estimation algorithm is an important element in the structure of the Shor factorisation algorithm. The usual version of the algorithm begins with a set of Hadamard operators in the circuit. It is to be noted that the Hadamard operation takes the computational basis (eigenbasis of the Pauli-z operator) to the eigenbasis of the Pauli-x operator, the elements of which are maximally coherent with respect to the conventional quantum coherence measures with the chosen basis being the computational one. We looked at the response of the efficiency of the quantum phase estimation algorithm if we replace the Hadamard matrices in its structure with non-Hadamard ones. Without loss of generality, we can characterise the non-Hadamard operations by the latitude of the pair of basis vectors on the Bloch sphere to which the operator takes the computational basis to. We found that the efficiency of non-Hadamard quantum phase estimation algorithms changes its behaviour again at a latitude that is approximately \(\frac{3 \pi}{10}\). More precisely,  it seems that while a maximal value of the conventional quantum coherence indicates the maximal efficiency of the quantum phase estimation algorithm, a maximal value of the noncommutative (quantum) coherence plays a complementary role by signalling the onset of a steep increase in the efficiency of the algorithm, while we sweep over non-Hadamard versions of the algorithm.


\end{document}